\documentclass[aps,12pt,preprintnumbers,nofootinbib,superscriptaddress]{revtex4}
\usepackage{graphicx}
\usepackage{amssymb,amsmath}
\begin{document}

\title{\Large{\bf Joule-Thomson expansion of charged Gauss-Bonnet black holes in AdS space } }

\author{Shan-Quan Lan}
\email[ ]{shanquanlan@126.com}
\affiliation{%
Department of Physics, Lingnan Normal University, Zhanjiang, 524048, Guangdong, China}

\date{\today}

\begin{abstract}
Joule-Thomson expansion process is studied for charged Gauss-Bonnet black holes in AdS space. Firstly, in five-dimensional space-time, the isenthalpic curve in $T-P$ graph is obtained and the cooling-heating region is determined. Secondly, the explicit expression of Joule-Thomson coefficient is obtained from the basic formulas of enthalpy and temperature. Our methods can also be applied to van der Waals system as well as other black hole systems. And the inversion curve $\tilde{T}(\tilde{P})$ which separates the cooling region and heating region is obtained and investigated. Thirdly, interesting dependence of the inversion curves on the charge $(Q)$ and the Gauss-Bonnet parameter $(\alpha)$ is revealed. In $\tilde{T}-\tilde{P}$ graph, the cooling region decreases with charge, but increases with the Gauss-Bonnet parameter. Fourthly, by applying our methods, the Joule-Thomson expansion process for $\alpha=0$ case in four dimension is studied, where the Gauss-Bonnet AdS black hole degenerates into RN-AdS black hole. The inversion curves for van der Waals systems consist of two parts. One has positive slope, while the other has negative slope. However, for black hole systems, the slopes of the inversion curves are always positive, which seems to be a universal feature.

\end{abstract}
\pacs{}
\maketitle
\newpage

\date{\today}

\section{Introduction}

Thermodynamics of black holes is an interesting and challenging topic since the discovery of black hole's entropy\cite{Bekenstein:1973ur}, the four thermodynamic law\cite{Bardeen1973}, and the Hawking radiation\cite{hawking1975} in 1970s. There exist fundamental connections between the classical thermodynamics, general relativity, and quantum mechanics. Specifically, due to the development of AdS/CFT duality\cite{Maldacena,Gubser,Witten}, these connections have been deepened and a lot attention has been attracted to the AdS black holes.

In AdS space, there exists Hawking-Page phase transition between stable large black hole and thermal gas\cite{Hawking} which is explained as the confinement/deconfinement phase transition of a gauge field\cite{Witten2}. Considering that the AdS black holes are electrically charged, rich phase structures are found by Chamblin et al\cite{Chamblin1,Chamblin2}. It is discovered that the phase transition behavior of charged AdS black hole is reminiscent to the liquid-gas phase transition in a van der Waals system\cite{Banerjee}. In the extended phase space where the cosmological constant is identified as pressure\cite{Kastor:2009wy}, further investigation of the $P-V$ critical behavior of a charged AdS black hole support the analogy between the black holes and the van der Waals liquid-gas system. It is found that the black hole and van der Waals system share not only the same $P-V$ diagram, but also the critical exponents\cite{2012JHEP07033K}. This analogy has been generalized to different AdS black holes, such as rotating black holes, higher dimensional black holes, Gauss-Bonnet black holes, f(R) black holes, black holes with scalar hair, etc\cite{Gunasekaran,Hendi,Chen,ZhaoZhao,Altamirano,Cai,AltamiranoKubiznak,XuXu,Mo,zou,MoLiu,Altamirano3,
Wei,Wei2,Zhang,Moliu,Zou2,Zhao2,ZhaoZhang,XuZhang,Frassino,Zhangcai,Mirza,Kostouki,Rajagopal,Liuwang,
Hendi:2017fxp,Hendi:2016vux,2013arXiv13053379S,Lan:2015bia,Ma:2017pap,Wei:2017icx,Bhattacharya:2017hfj,
Hendi:2016usw,Kuang:2016caz,Fernando:2016sps,Majhi:2016txt,Zeng:2016aly,Sadeghi:2016dvc,Zeng:2015wtt,
Nguyen:2015wfa,Xu:2015rfa,Nie:2015zia}.

Apart from the phase transition and critical phenomena, the analogy between the black holes and the van der Waals system has been creatively generalized to the well-known Joule-Thomson expansion process\cite{Aydiner1} recently. For Joule-Thomson expansion in classical thermodynamics, gas at a high pressure passes through a porous plug to a section with a low pressure and during the process enthalpy is unchanged. An interesting phenomenon during Joule-Thomson expansion process is that the $T-P$ graph is divided into two parts. One is cooling region, while the other is heating region. They are separated by the so-called inversion curve. For charged AdS black holes\cite{Aydiner1} and Kerr-AdS black holes\cite{Aydiner2}, the isenthalpy expansion process and the inversion curve are investigated. Then the works are generalized to quintessence charged AdS black holes\cite{Ghaffarnejad}, holographic superfluids\cite{Yogendran}, charged AdS black holes in f(R) gravity\cite{Chabab},  AdS black hole with a global monopole\cite{ahmedrizwan} and AdS black holes in Lovelock gravity\cite{moli1805}. The results show that the inversion curves $\tilde{T}(\tilde{P})$ for all the above black hole systems have only positive slope. While for the van der Waals system, the inversion curves have both positive and negative slopes which form a circle in the pressure axis.

We are curious about the missing of negative slopes of the inversion curves, and eager to check out that whether the quantum gravity effects can cure this problem or whether this feature for black hole systems is just universal. So we will focus on the Gauss-Bonnet Einstein-Maxwell gravity.  Considering the effects of Gauss-Bonnet term is interesting and important. Whatever the quantum gravity may be, there will be higher order corrections to the pure Einstein action, and Gauss-Bonnet is a well combination of terms which have candidate corrections. What's more, these terms represent part of the $1/N$ correction to the large $N$ limit of the holographically dual $SU(N)$-like gauge field theory\cite{oassgubser}. As a result, the investigation is interesting in its own right.

This paper is organized as follows. In Sec.\ref{sec2}, we briefly review the D-dimensional Charged Gauss-Bonnet black holes in AdS space. In Sec.\ref{sec3}, we investigate Joule-Thomson expansion for the Gauss-Bonnet parameter $\alpha>0$ and $D=5$ dimensional black holes. The isenthalpy curve is studied and two methods are introduced to derive an explicit Joule-Thomson coefficient. The effects of charge and Gauss-Bonnet parameter on the inversion curves are studied. And we also investigate Joule-Thomson expansion for the Gauss-Bonnet parameter $\alpha=0$ and $D=4$ dimensional black holes via our new method. Sec.\ref{sec4} devotes to conclusion and discussion.

\section{A brief review of the D-dimensional Charged Gauss-Bonnet black holes in AdS space}
\label{sec2}

Consider the action in D-dimensional Einstein-Maxwell theory with a cosmological constant $\Lambda$ and a Gauss-Bonnet term:
\begin{eqnarray}
  S=\frac{1}{16\pi}\int d^{D}x\sqrt{-g}[R-2\Lambda+\alpha_{GB}(R_{\gamma\delta\mu\nu}R^{\gamma\delta\mu\nu}-4R_{\mu\nu}R^{\mu\nu}+R^{2})-F^{2}],
\end{eqnarray}
where the Gauss-Bonnet parameter $\alpha_{GB}$ has dimesions of $[length]^{2}$ and the cosmological constant $\Lambda=-\frac{(D-1)(D-2)}{2l^{2}}$. When $\alpha_{GB}=0$, the action degenerates into Einstein-Maxwell theory in AdS space. When $\alpha_{GB}\neq0$, we will work in $D\geq5$, since in the case $D=4$ the Gauss-Bonnet term is purely topological. In this paper, we define $\alpha\equiv (D-3)(D-4)\alpha_{GB}$.

The action admits a static black hole solution with maximal symmetry as
\begin{eqnarray}
  ds^{2}=-Y(r)dt^{2}+\frac{dr^{2}}{Y(r)}+r^{2}d\Omega_{D-2}^{2},
\end{eqnarray}
where $d\Omega_{D-2}^{2}$ is the metric on a round $D-2$ sphere with volume $\omega_{D-2}$, and
\begin{eqnarray}\label{eqp}
  Y(r)=1+\frac{r^{2}}{2\alpha}(1-\sqrt{1+\frac{4\alpha m}{r^{D-1}}-\frac{4\alpha q^{2}}{r^{2D-4}}-\frac{4\alpha}{l^{2}}}).
\end{eqnarray}
Notice that since the $m=q=0$ case defines the vacuum solution for a given value of $l$, $\alpha$ cannot be arbitrary\cite{boulwaresd}, but must be constrained by $0\leq 4\alpha/l^{2}\leq 1$. A study on the quasi-normal modes of the $D=5$ black holes show that there is eikonal instability for $\alpha/l^{2}\geq 0.8$\cite{rakonoplya1,rakonoplya2}. So we should adopt small $\alpha$ and carefully consider the constrains in the following investigation.

When $\alpha=0$, using the L'Hospital's rule, Eq.(\ref{eqp}) will become
\begin{eqnarray}
  Y(r)=1-\frac{m}{r^{D-3}}+\frac{q^{2}}{r^{2D-6}}+\frac{r^{2}}{l^{2}},
\end{eqnarray}
which is the familiar charged AdS black hole. $m,q$ are related to the ADM mass $M$, electric charge Q by
\begin{eqnarray}
  M=\frac{(D-2)\omega_{D-2}}{16\pi}m,\,\,\,\,\,\,Q=\sqrt{2(D-2)(D-3)}\frac{\omega_{D-2}}{8\pi}q.
\end{eqnarray}

The horizon radius $r_{+}$ of the black hole is the largest root of $Y(r_{+})=0$,which gives us an equation for the black hole mass $M$,
\begin{eqnarray}\label{admmass}
  M=\frac{(D-2)\omega_{D-2}}{16\pi}(\alpha r_{+}^{D-5}+r_{+}^{D-3}+\frac{q^{2}}{r_{+}^{D-3}}+\frac{16\pi P r_{+}^{D-1}}{(D-1)(D-2)}).
\end{eqnarray}
Note that the AdS radius $l$ is replaced by pressure $P=\frac{(D-1)(D-2)}{16\pi l^{2}}$, and in this extended phase space the black hole mass is treated as enthalpy instead of internal energy. The temperature is obtained by the first derivative of $Y(r)$ at the horizon,
\begin{eqnarray}\label{temperature}
  T=\frac{Y'(r_{+})}{4\pi}=\frac{1}{4\pi r_{+}(r_{+}^{2}+2\alpha)}(\frac{16\pi P r_{+}^{4}}{D-2}+(D-3)r_{+}^{2}+(D-5)\alpha-(D-3)\frac{q^{2}}{r_{+}^{2D-8}}).
\end{eqnarray}

In the following, we will only use Eq.(\ref{admmass}) and Eq.(\ref{temperature}) to find the isenthalpic curves and determine the inversion temperatures between the cooling and heating regions for the black hole system in $T-P$ plane.

\section{Joule-Thomson Expansion}
\label{sec3}

The Joule-Thomson expansion for black hole system is an isenthalpy process in the extended phase space, the enthalpy $H$ is identified as the black hole mass $H=M$. Similar to the Joule-Thomson process with fixed particle number for van der Waals gases, we should consider the canonical ensemble with fixed charge $Q$. The Gauss-Bonnet parameter $\alpha$ will also be treated as a constant.

\subsection{ $\alpha> 0$ and $D=5$}

\begin{figure}
		\centering
		\includegraphics[scale=0.6]{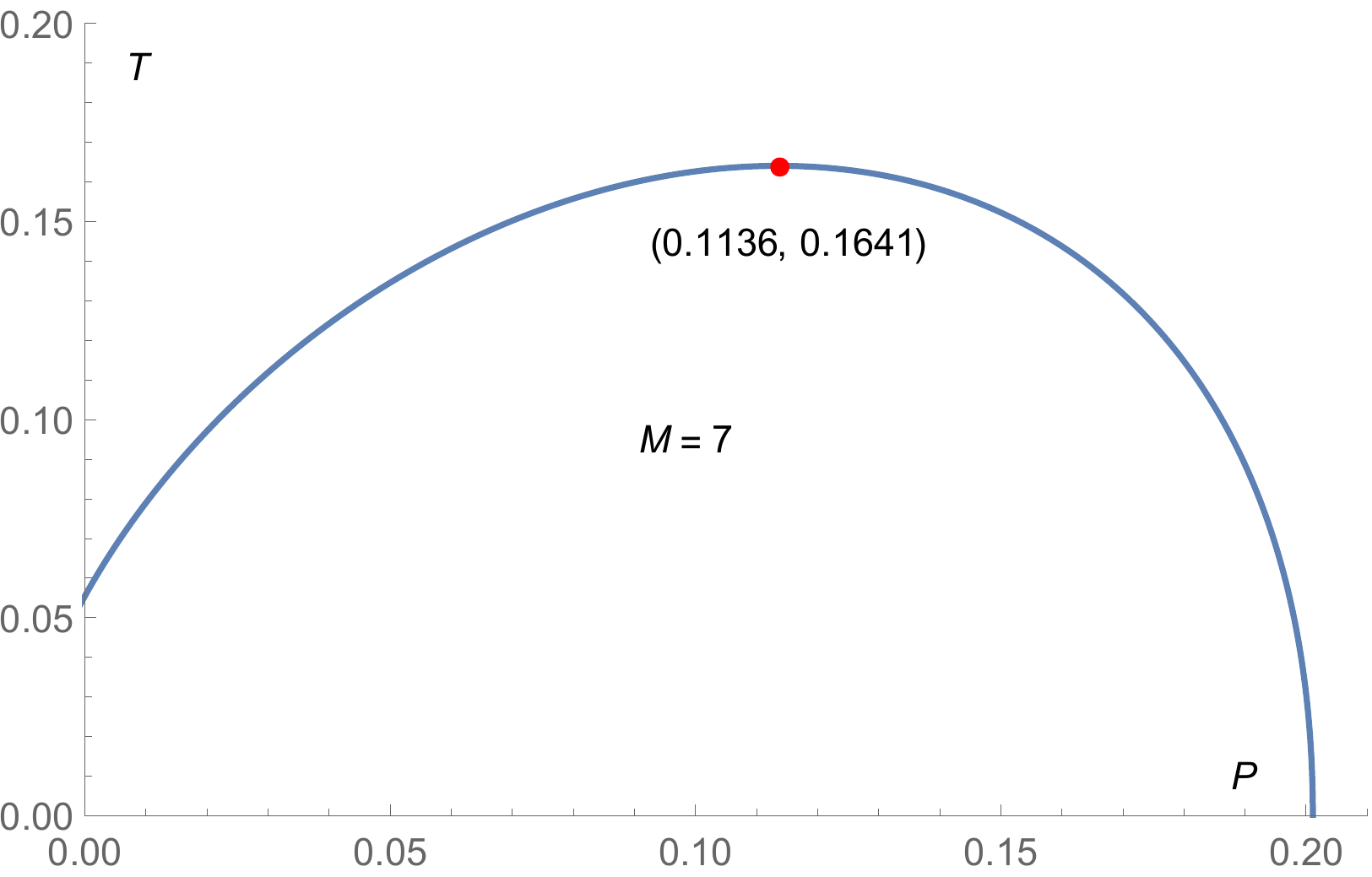}
	\caption{The isenthalpy process  with $M=7$ in $T-P$ graph at $\alpha=0.2$, $Q=\sqrt{3}\pi$. The inversion pressure and temperature between the cooling and heating regions is $(\tilde{P}=0.1136,\tilde{T}=0.1641)$ where the Joule-Thomson coefficient $\mu=(\frac{\partial T}{\partial P})_{M}=0$.}
	\label{tpm1}
\end{figure}

When $\alpha> 0$ and $D=5$, Eq.(\ref{admmass}) and Eq.(\ref{temperature}) become
\begin{eqnarray}\label{tprrpm}
  M&=&\frac{3\pi}{8}(\alpha +r_{+}^{2}+\frac{4Q^{2}}{3\pi^{2}r_{+}^{2}}+\frac{4}{3}\pi P r_{+}^{4}),\nonumber\\
  T&=&\frac{\pi^{2}r_{+}^{4}(3+8\pi P r_{+}^{2})-4Q^{2}}{6\pi^{3}r_{+}^{3}(2\alpha+r_{+}^{2})}.
\end{eqnarray}
The pressure $P$ can be rewritten as a function of $M$ and $r_{+}$, then substitute $P(M,r_{+})$ into the temperature formula which also become a function of $M$ and $r_{+}$,
\begin{eqnarray}\label{tpmr}
  P(M,r_{+})&=&\frac{8\pi M r_{+}^{2}-3\pi^{2}\alpha r_{+}^{2}-3\pi^{2}r_{+}^{4}-4Q^{2}}{4\pi^{3}r_{+}^{6}},\nonumber\\
  T(M,r_{+})&=&\frac{16\pi M r_{+}^{2}-6\pi^{2}\alpha r_{+}^{2}-3\pi^{2}r_{+}^{4}-12Q^{2}}{6\pi^{3}r_{+}^{3}(2\alpha+r_{+}^{2})}.
\end{eqnarray}
Solving the above equations, one can obtain function $T(M,P)$ which is lengthy and won't be shown here. For particular values of $\alpha=0.2$, $Q=\sqrt{3}\pi$, the $T(M,P)$ curve is shown for $M=7$ in Fig.\ref{tpm1}. According to the definition of Joule-Thomson coefficient $\mu=(\frac{\partial T}{\partial P})_{M}$, the inversion pressure and temperature between the cooling and heating regions is $(\tilde{P}=0.1136,\tilde{T}=0.1641)$ which is set by $\mu=0$. So the most important thing is to find the function expression of $\mu$. By setting $\mu=0$, one can obtain the inversion points $(\tilde{P},\tilde{T})$ for different fixed enthalpy $M$.

In Ref.\cite{Aydiner1}, the Joule-Thomson coefficient is given by,
\begin{eqnarray}
  \mu=(\frac{\partial T}{\partial P})_{M}=\frac{1}{C_{P}}[T(\frac{\partial V}{\partial T})_{P}-V],
\end{eqnarray}
which is elegant. But in this paper, we will adopt more straightforward methods by using only Eq.(\ref{admmass}) and Eq.(\ref{temperature}) to derive the Joule-Thomson coefficient $\mu$. From Eq.(\ref{tprrpm}), one can find that temperature is a function of pressure and radius, radius is a function of pressure and mass. Then the Joule-Thomson coefficient is given by,
\begin{eqnarray}
  \mu&=&(\frac{\partial T}{\partial P})_{M}=(\frac{\partial T}{\partial P})_{r_{+}}+(\frac{\partial T}{\partial r_{+}})_{P}(\frac{\partial r_{+}}{\partial P})_{M}=(\frac{\partial T}{\partial P})_{r_{+}}+(\frac{\partial T}{\partial r_{+}})_{P}/(\frac{\partial P}{\partial r_{+}})_{M}\nonumber\\
  &=&\frac{4r_{+}^{3}}{3(2\alpha+r_{+}^{2})}+
  \frac{r_{+}^{3}(20 Q^{2}r_{+}^{2}+\pi^{2} r_{+}^{6}(8\pi P r_{+}^{2}-3)+6\alpha(4Q^{2}+\pi^{2}r_{+}^{4}(1+8\pi P r_{+}^{2})))}{3(2\alpha+r_{+}^{2})^{2}(12Q^{2}+\pi r_{+}^{2}(3\pi(2\alpha+r_{+}^{2})-16M))}.
\end{eqnarray}
From Eq.(\ref{tpmr}), one can obtain a more simple expression,
\begin{eqnarray}
  \mu&=&(\frac{\partial T}{\partial P})_{M}=(\frac{\partial T}{\partial r_{+}})_{M}(\frac{\partial r_{+}}{\partial P})_{M}=(\frac{\partial T}{\partial r_{+}})_{M}/(\frac{\partial P}{\partial r_{+}})_{M}\nonumber\\
  &=&\frac{r_{+}^{3}(3\pi^{2}r_{+}^{6}+12\pi(\pi\alpha -4M)r_{+}^{4}+4(\pi\alpha(3\pi\alpha-8M)+15Q^{2})r_{+}^{2}+72\alpha Q^{2})}{3(2\alpha+r_{+}^{2})^{2}(12Q^{2}+\pi r_{+}^{2}(3\pi(2\alpha+r_{+}^{2})-16M))}.
\end{eqnarray}

\begin{figure}
		\centering
		\includegraphics[scale=0.6]{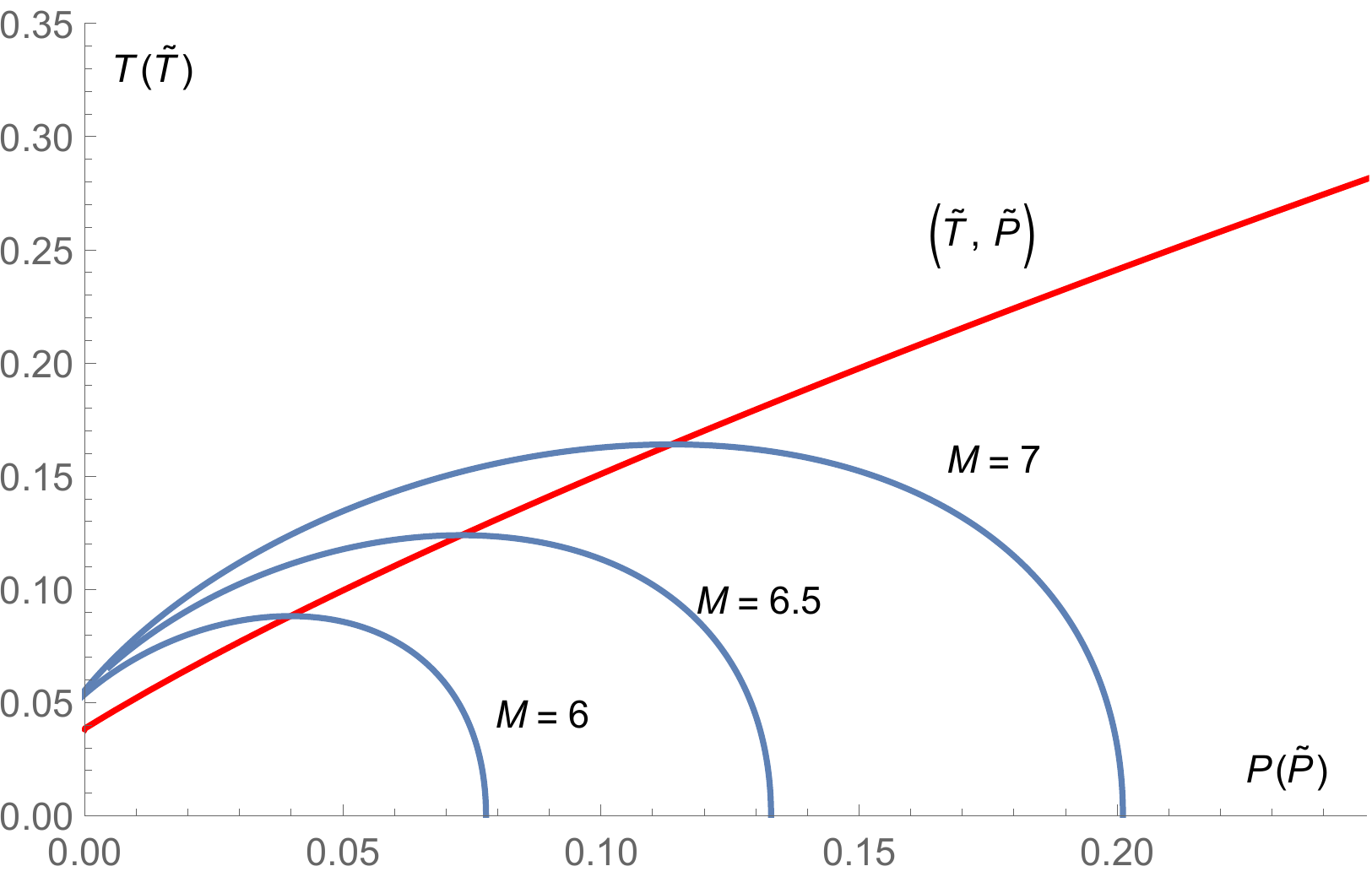}
	\caption{$T$ vs. $P$ and $\tilde{T}$ vs. $\tilde{P}$ for $\alpha=0.2$, $Q=\sqrt{3}\pi$. The blue ones are isenthalpy curves $T(M,P)$ for $M=6,6.5,7$. The red one is the inversion curve  $\tilde{T}(\tilde{P})$. One can find that the inversion points $(\tilde{P},\tilde{T})$ increase monotonically with enthalpy.}
	\label{tpm2}
\end{figure}

All the above methods are equivalent. Choose one method and set $\mu=0$, together with Eq.(\ref{tprrpm}) or Eq.(\ref{tpmr}), one can obtain the inversion points $(\tilde{P},\tilde{T})$. For our Gauss-Bonnet case, the final result is very lengthy. So we will only show the inversion curve in Fig.\ref{tpm2} where $\alpha=0.2$, $Q=\sqrt{3}\pi$. The blue ones are isenthalpy curves $T(M,P)$ for different enthalpy $M$. The red one is the inversion curve  $\tilde{T}(\tilde{P})$. One can find that the inversion points $(\tilde{P},\tilde{T})$ increase monotonically with enthalpy, suggesting that the inversion temperature and pressure are higher for the Joule-Thomson expansion with a larger enthalpy. The inversion curve divides each isenthalpic curve into two branches. The branch above the inversion curve, which always have positive slope, represents the cooling process. And the branch below the inversion curve, which always have negative slope, represents the heating process.

\begin{figure}
		\centering
		\includegraphics[scale=0.45]{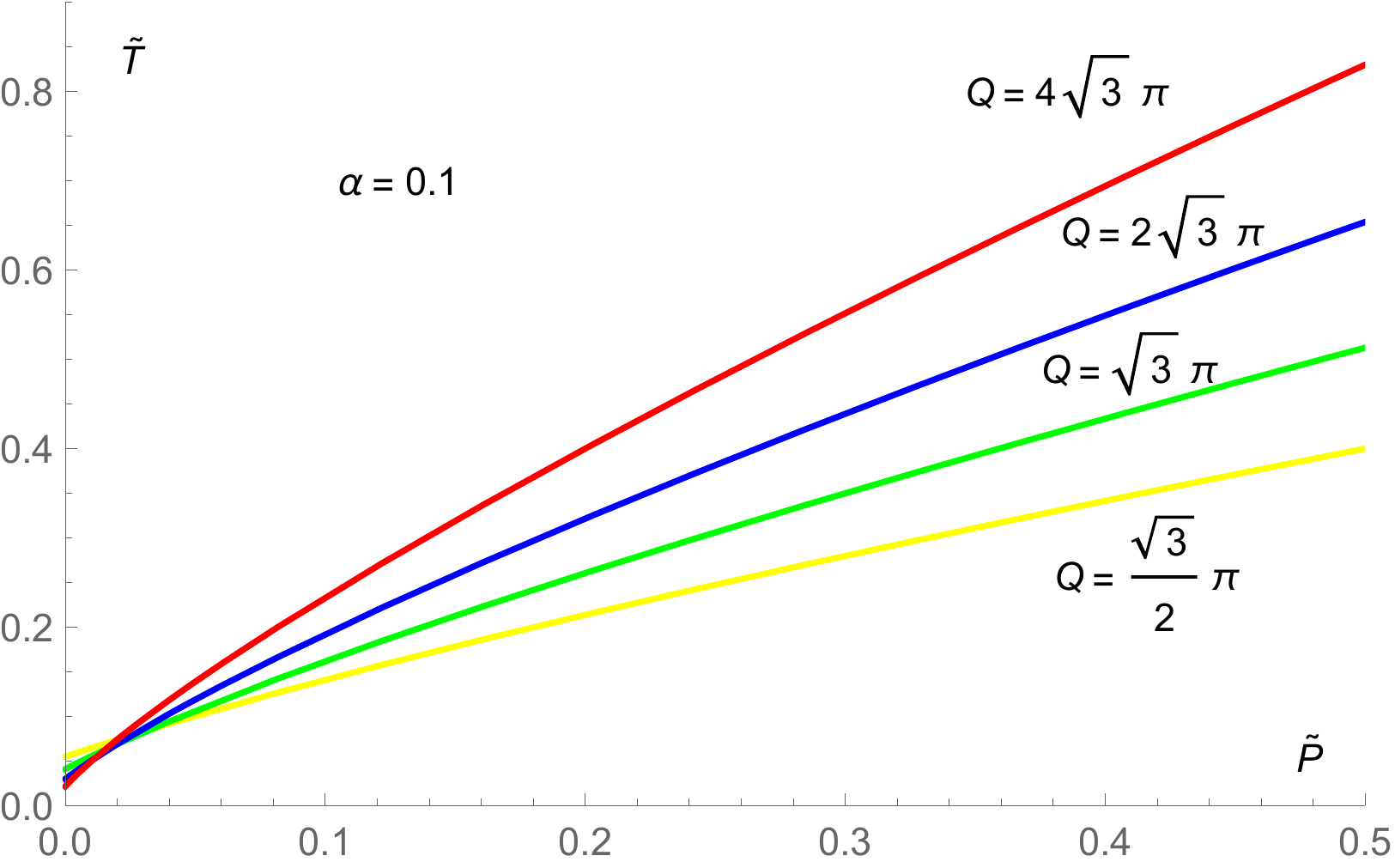}
        \includegraphics[scale=0.45]{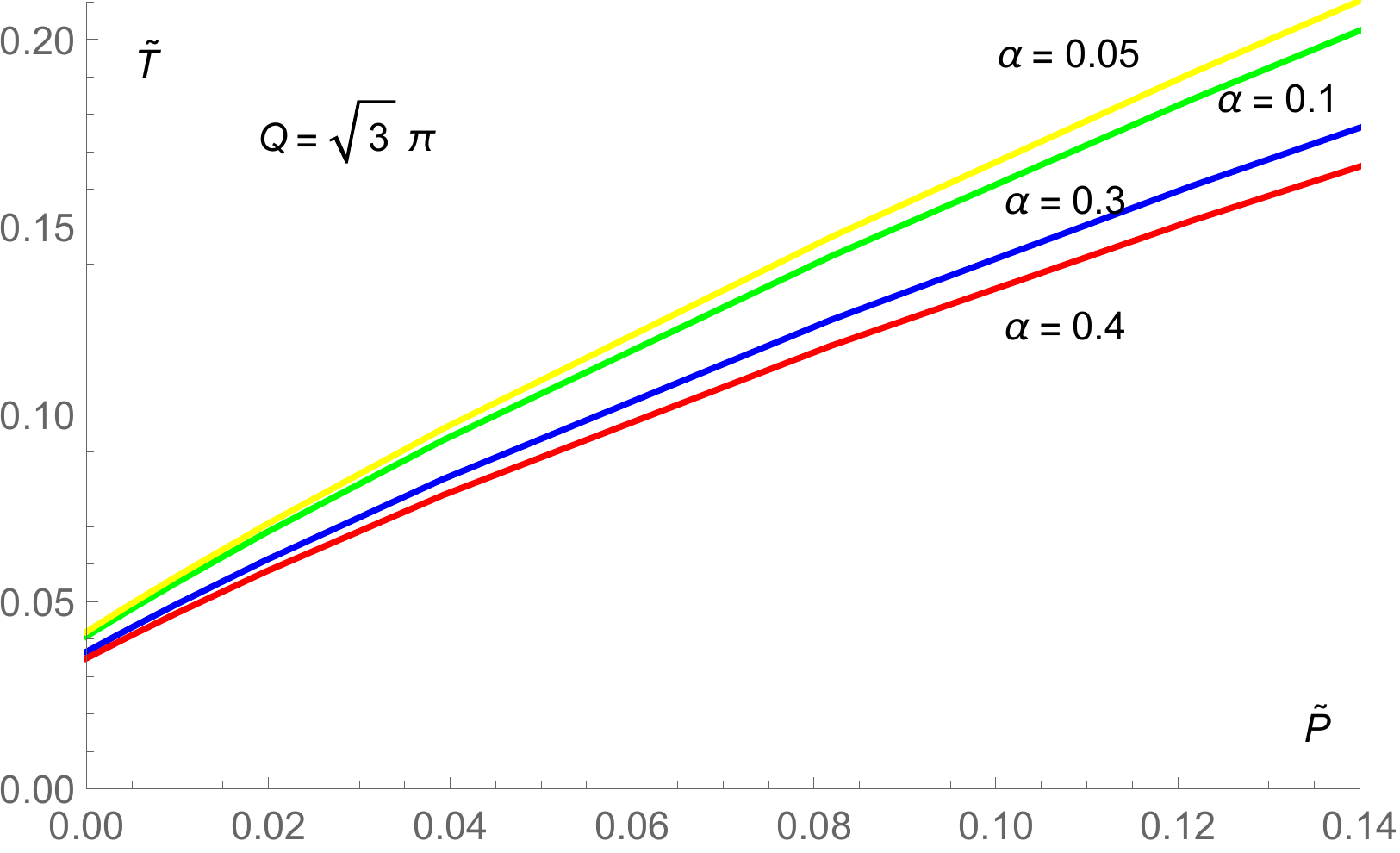}
	\caption{The dependence of the inversion curves on the charge and the Gauss-Bonnet parameter.  Left panel is the inversion curves for charge $Q=\frac{\sqrt{3}}{2}\pi, \sqrt{3}\pi,  2\sqrt{3}\pi,4\sqrt{3}\pi$ with fixed Gauss-Bonnet parameter $\alpha=0.1$. The slope of the inversion curves increase with charge. Right panel is the inversion curves for Gauss-Bonnet parameter $\alpha=0.05, 0.1, 0.3, 0.4$ with fixed charge $Q=\sqrt{3}\pi$. The slope of the inversion curves decrease with Gauss-Bonnet parameter.}
	\label{tpaq}
\end{figure}
It is of interest to probe the dependence of the inversion points on the charge and the Gauss-Bonnet parameter.  Left panel of Fig.\ref{tpaq} shows the relationship between the inversion curves and the charge for fixed Gauss-Bonnet parameter $\alpha=0.1$. The slope of the inversion curves increase with the charge, suggesting that the inversion temperature is higher for the Joule-Thomson expansion with a larger charge. Right panel of Fig.\ref{tpaq} shows the relationship between the inversion curves and Gauss-Bonnet parameter for fixed charge $Q=\sqrt{3}\pi$. In contrast to the effect of charge, the slope of the inversion curves decrease with Gauss-Bonnet parameter, suggesting that the inversion temperature is lower for the Joule-Thomson expansion with a larger Gauss-Bonnet parameter. So the effect of charge and Gauss-Bonnet parameter are different. In $\tilde{T}-\tilde{P}$ graph, the cooling region decreases with charge, but increases with the Gauss-Bonnet parameter.


\subsection{ $\alpha=0$ and $D=4$}

When $\alpha= 0$ and $D=4$, the Gauss-Bonnet AdS black hole degenerates into RN-AdS black hole. Its Joule-Thomson expansion process is investigated in Ref.\cite{Aydiner1}. In this section, we will use our methods to reinvestigate the Joule-Thomson expansion and  double check these methods.

The starting point is still Eq.(\ref{admmass}) and Eq.(\ref{temperature}) which become
\begin{eqnarray}
  M&=&\frac{3Q^{2}+3r_{+}^{2}+8\pi P r_{+}^{4}}{6r_{+}},\nonumber\\
  T&=&\frac{-Q^{2}+r_{+}^{2}+8\pi P r_{+}^{4}}{4\pi r_{+}^{3}},
\end{eqnarray}
or rewritten as
\begin{eqnarray}
  P(M,r_{+})&=&\frac{6M r_{+}-3Q^{2}-3r_{+}^{2}}{8\pi r_{+}^{4}},\nonumber\\
  T(M,r_{+})&=&\frac{3M r_{+}-2Q^{2}-r_{+}^{2}}{2\pi r_{+}^{3}}.
\end{eqnarray}
Then the Joule-Thomson coefficient is obtained as,
\begin{eqnarray}
  \mu&=&(\frac{\partial T}{\partial P})_{r_{+}}+(\frac{\partial T}{\partial r_{+}})_{P}/(\frac{\partial P}{\partial r_{+}})_{M}\nonumber\\
  &=&\frac{r_{+}(15Q^{2}+r_{+}(-18M+5r_{+}+8\pi P r_{+}^{3}))}{3(2Q^{2}+r_{+}(-3M+r_{+}))}\nonumber\\
  &=&\frac{4r_{+}(-3Q^{2}+2r_{+}^{2}+8\pi P r_{+}^{4})}{3(-Q^{2}+r_{+}^{2}+8\pi P r_{+}^{4})},
\end{eqnarray}
or
\begin{eqnarray}
  \mu&=&(\frac{\partial T}{\partial r_{+}})_{M}/(\frac{\partial P}{\partial r_{+}})_{M}\nonumber\\
  &=&\frac{2r_{+}(6Q^{2}+r_{+}(r_{+}-6M))}{3(2Q^{2}+r_{+}(r_{+}-3M))}\nonumber\\
  &=&\frac{4r_{+}(-3Q^{2}+2r_{+}^{2}+8\pi P r_{+}^{4})}{3(-Q^{2}+r_{+}^{2}+8\pi P r_{+}^{4})}.
\end{eqnarray}
They are consistent with each other. Setting $\mu=0$, the only positive and real root of $r_{+}$ is
\begin{equation}
  r_{+}=\frac{\sqrt{\sqrt{1+24\pi  Q^{2}\tilde{P}}-1}}{2\sqrt{2\pi \tilde{P}}}.
\end{equation}
Substituting this root into the temperature formula above, finally the analytical inversion curve $\tilde{T}(\tilde{P})$ is obtained
\begin{eqnarray}
  \tilde{T}=\sqrt{\frac{\tilde{P}}{2\pi}}\frac{(1+16\pi Q^{2}\tilde{P}-\sqrt{1+24\pi Q^{2}\tilde{P}})}{(\sqrt{1+24\pi  Q^{2}\tilde{P}}-1)^{3/2}},
\end{eqnarray}
which is exactly the same as Eq.(44) in Ref.\cite{Aydiner1}. Similar to $\alpha>0, D=5$ case, the slopes of the inversion curves $\tilde{T}(\tilde{P})$ are also always positive.

\section{Conclusion and discussion}
\label{sec4}

In this paper, we studied the Joule-Thomson expansion for charged Gauss-Bonnet black holes in AdS space.  In the extended phase, the cosmological is identified as pressure while the black hole mass is identified as enthalpy. Thus we considered the expansion process with constant mass. Firstly, the Joule-Thomson expansion process (the isenthalpy curve) is depicted in Fig.\ref{tpm1}. The inversion point $(\tilde{P},\tilde{T})$ separating the cooling region and heating region locates at $\mu=0$ where the temperature is highest during the Joule-Thomson expansion process. Left part with positive slope is the cooling region, while right part with negative slope is the heating region.

Secondly, the explicit expression of Joule-Thomson coefficient is obtained from general formulas of enthalpy and temperature. Our methods can also be applied to van der Waals gas as well as other black hole systems. The isenthalpy curves for different enthalpy and the inversion curve for specific Gauss-Bonnet parameter $\alpha=0.2$ and charge $Q=\sqrt{3}\pi$ are depicted in Fig.\ref{tpm2}. The inversion curve divides the $T-P$ graph into two branches. The branch above the inversion curve is the cooling region, and the branch below the inversion curve is the heating region. The slope of the inversion curve is always positive.

Thirdly, the dependence of the inversion curves on the charge and the Gauss-Bonnet parameter is investigated. An interesting result is depicted in Fig.\ref{tpaq} where the slope of the inversion curves increase with charge, but decrease with Gauss-Bonnet parameter. As a result, the effect of the charge and the Gauss-Bonnet parameter are different. In $\tilde{T}-\tilde{P}$ graph, the cooling region decreases with charge, but increases with the Gauss-Bonnet parameter.

Finally, we checked that the Gauss-Bonnet AdS black hole degenerates into RN-AdS black hole when $\alpha=0$. By applying our methods, the Joule-Thomson expansion process is reinvestigated in four dimensional space-time. An analytical expression of the inversion curves is obtained, which is exactly same as that in Ref.\cite{Aydiner1}. The slope of the inversion curves are also always positive.

Recently, in Ref.\cite{moli1805}, the Joule-Thomson expansion of Lovelock--AdS black hole is studied from various perspectives. Their paper and this paper focus on the same aspect of black holes. Namely, the Joule-Thomson expansion. However, there are no overlaps between the two papers, even considering that the Gauss-Bonnet--AdS black hole is a particular case of the Lovelock--AdS black hole. Their paper only discussed a specific Lovelock solution where the Gauss-Bonnet Lagrangian and the third Lovelock Lagrangian are connected by one parameter. Moreover, the dimensionality of the space-time being considered is different. They focus on seven-dimensional space-time while we studied the five-dimensional Gauss-Bonnet black hole. As the low-energy limit of string theory, the Gauss-Bonnet gravity and its black hole solution is of physical significance in its own right. What's more, one important contribution of the paper is that we propose new methods to calculate the Joule-Thomson coefficient which is a key quantity in this topic.

In the end, we would like to point out that the inversion curves of Joule-Thomson expansion for van der Waals gas system consist of two parts which forms a circle in pressure axis. One part is a lower one with positive slope, the other is an upper one with negative slope. For the details, one can refer to Ref.\cite{Aydiner1}, or any textbook on thermodynamics and statistical physics. But as far as we know, the inversion curves of Joule-Thomson expansion for black hole system, such as charged AdS black holes\cite{Aydiner1}, Kerr-AdS black holes\cite{Aydiner2}, quintessence charged AdS black holes\cite{Ghaffarnejad}, charged AdS black holes in f(R) gravity\cite{Chabab}, AdS black hole with a global monopole\cite{ahmedrizwan}, AdS black holes in Lovelock gravity\cite{moli1805}, as well as our charged AdS black holes in Gauss-Bonnet gravity, consist only positive slope branch.  This seems to be an universal feature for black hole systems. The underlying physics behind the missing of the negative slope or the difference of Joule-Thomson expansion between black hole system and van der Waals system deserves to be disclosed in the future research.

\begin{acknowledgments}
This research is supported by Department of Education of Guangdong Province, China (Grant Nos.2017KQNCX124).
\end{acknowledgments}


\end{document}